\documentclass[10pt,letterpaper,twocolumn]{article} 

\usepackage{ol2}
\usepackage{balance}
\usepackage[draft,implicit=false]{hyperref}
\usepackage{amsmath}
\usepackage{filecontents}

\begin{document}

\twocolumn[

\title{Observing spin Hall effect of pseudo-thermal light through weak measurement}


\author{Bin Cao,$^{1,2}$ Dong Wei,$^{1,3}$ Pei Zhang,$^1$ Hong Gao,$^{1,*}$ and Fuli Li$^{1}$}

\address{
$^1$Department of Applied Physics, Xi'an Jiaotong University, Xi'an  710049, China\\
$^2$Joint Quantum Institute, NIST/University of Maryland, College Park, Maryland 20742, USA\\
$^3$e-mail: weidong@mail.xjtu.edu.cn\\
$^*$Corresponding author: honggao@mail.xjtu.edu.cn
}

\begin{abstract}
It is well known that the spin Hall effect of light (SHEL) can be easily observed via weak measurement by the dark strip resulted from the splitting of different polarization components. We find that the SHEL of partially coherent beam (PCB) also has similar phenomenon. However, the existence of the dark strip in the SHEL of PCB can not be properly justified by considering the beam as a statistical assemble of coherent speckles. Also, the dark strip of PCB is not purely dark. By analyzing the autocorrelation of the central dark strip part of the SHEL of PCB, we show that SHEL of PCB is essentially the direct result of overlapping coherent speckles' SHEL. Then we further prove our conclusion by adjusting the converging level and the incident angle of the beam. Finally, we develop a qualitative theory by taking the intensity distribution of PCB into account, to fully clarify the SHEL of PCB. The feasibility of our theory to Goos-H\"anchen (GH) scenario is also demonstrated.
\end{abstract}

\ocis{030.1640, 240.3695, 070.0070.}

 ] 

The spin Hall effect of light (SHEL) \cite{fedorov1955to} has attracted growing attention as a result of the rapid development of optics at nano and subwavelength
scales \cite{Overview,onoda2004hall}. SHEL and the Goos-H\"anchen shift \cite{goos1955TEin}, which are both caused by the conservation of photon's momentum, have been explored in different areas \cite{jayaswal2014observation,jayaswal2013weak,wang2013goos,chuang2015negative,rodriguez2010optical,Ziauddin2012control,haefner2009spin}. The research of the SHEL of complex incident beams and the SHEL at special incident angles bring about brand-new analysis of reflection/refraction \cite{korger2014observation,lv2012spin} and the new experimental technic: weak measurement \cite{Science,dressel2014colloquium}. Besides, the applications of SHEL also produce new methods to examine the optical properties of complex media \cite{luo2009spin,nalitov2015spin}.

The spin Hall effect of pseudo-thermal light has been predicted theoretically \cite{Aiello_0} and confirmed experimentally \cite{Aiello_1,merano2012observation}. It is shown that partially coherent light exhibits the same property with coherent beam in spatial shifts, but behaves differently in angular shifts.  Different from \cite{Aiello_1,merano2012observation}, in this paper we use weak measurement to observe the SHEL of pseudo-thermal light. Weak measurement \cite{Science,dressel2014colloquium} is useful in the amplification and detection of weak effects. It amplifies the SHEL of each single speckle of the pseudo-thermal light, making detection and processing easier. 

By using the existing theory \cite{Science}, we can firstly predict the SHEL of pseudo-thermal light. Coherent beams are transferred to pseudo-thermal light with rotating frosted glass. The roughness decides the spatial coherence of the output beam and the rate of rotation affects the temporal coherence. For highly coherent beam, the visual appearance of SHEL via weak measurement is the dark strip on the observation plane resulted from the splitting of right and left circular polarization light \cite{Measurement}. Therefore the experimental appearance of SHEL of pseudo-thermal light should merely be a homogenous darkening of the intensity distribution if we regard pseudo-thermal light beam as assemble of small coherent Gaussian speckles. In contrast, our results are not simple as our expectation. The result of the SHEL of PCB also shows a dark strip in the center of the beam, which is similar to the SHEL of coherent beam.

\begin{figure}[htbp]
\centerline{\includegraphics[width=1\columnwidth]{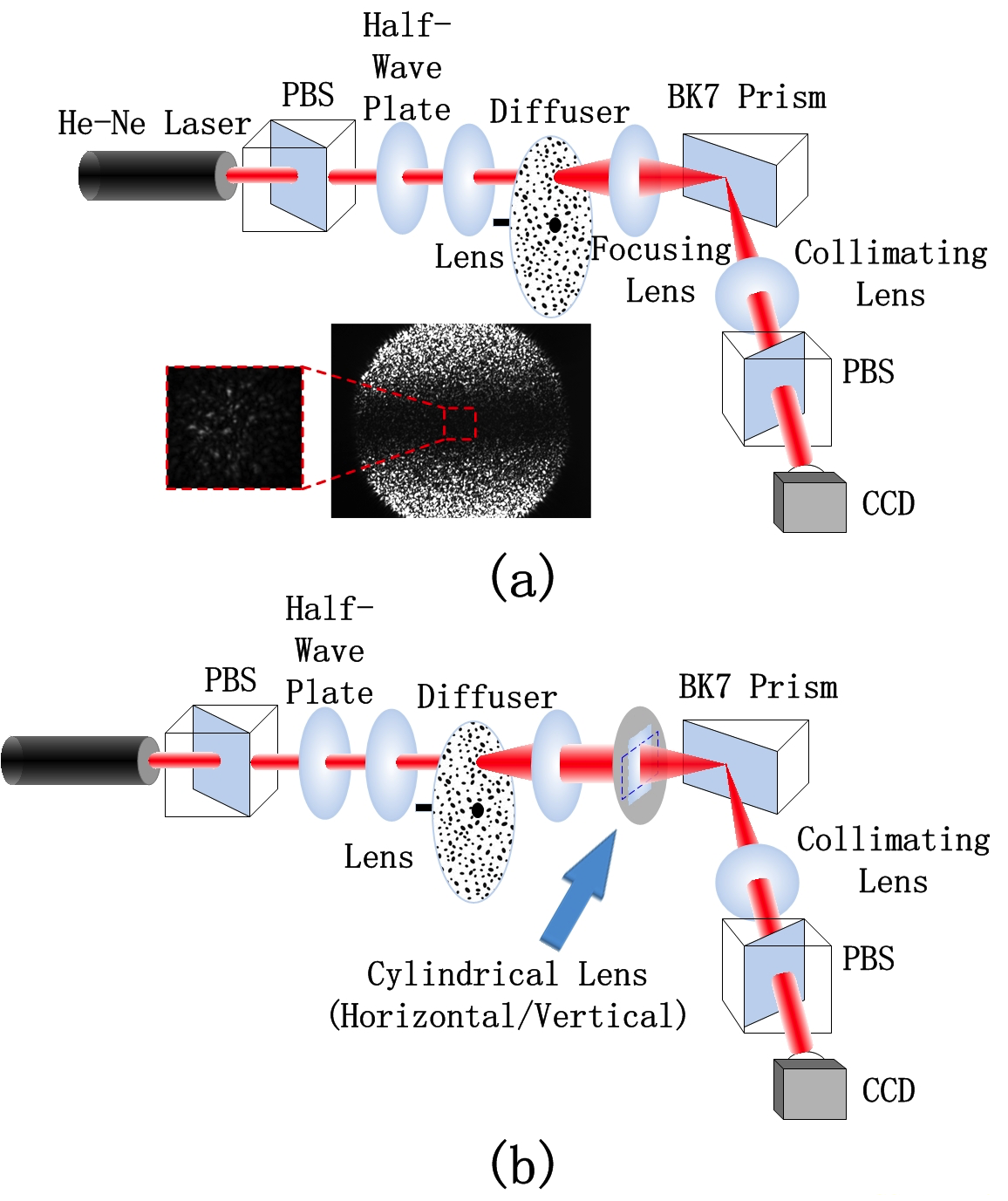}}
\caption{(a) Setup used to observe the SHEL of pseudo-thermal light via weak measurement. The diffuser is kept still when taking the inset picture. The incident angle is set to be $48.3^{\circ}$. (b) Setup to test the effect of overlapping density. A cylindrical lens is added to vary the overlapping density of unit speckles. When the cylindrical lens is set horizontal (vertical), the superposition density is relatively large (small), thus the dark strip of pseudo-thermal light is expected to be narrow (broad). Results captured by CCD are given in Fig. \ref{2}. The polarizer is deviated by $0.3^{\circ}$ from being crossed with the incident beam's polarization, which decides the amplification factor of weak measurement.}
\label{1}
\end{figure}

We use setup shown as Fig. \ref{1}(a) for the observation of SHEL of pseudo-thermal light via weak measurement. The laser beam of 632.8 nm is firstly filtered by a polarized beam splitter (PBS) in order to purify the incident beam as linear polarized. By adding a half-wave plate, we get the ability to change the direction of polarization. After a set of Lens-Diffuser-Lens, the beam is transferred to pseudo-thermal light and focused onto the surface of the glass prism. Due to the conservation of photon's momentum in reflection, the SHEL of PCB happens on the surface. The beam is then collimated and pass through a second PBS, which accomplishes the weak measurement process. Afterwards, we acquire the result by the CCD. The intensity distribution captured appears a dark strip, as given in the inset of Fig. \ref{1}(a). However, the dark strip isn't purely dark. Even though the average intensity of the strip is lower than the adjacent parts, there are speckles spreading in this area. As mentioned above, this result can't be fully explained if regarding the pseudo-thermal light as a simple assemble of coherent speckles.

\begin{figure}[htbp]
\centerline{\includegraphics[width=.8\columnwidth]{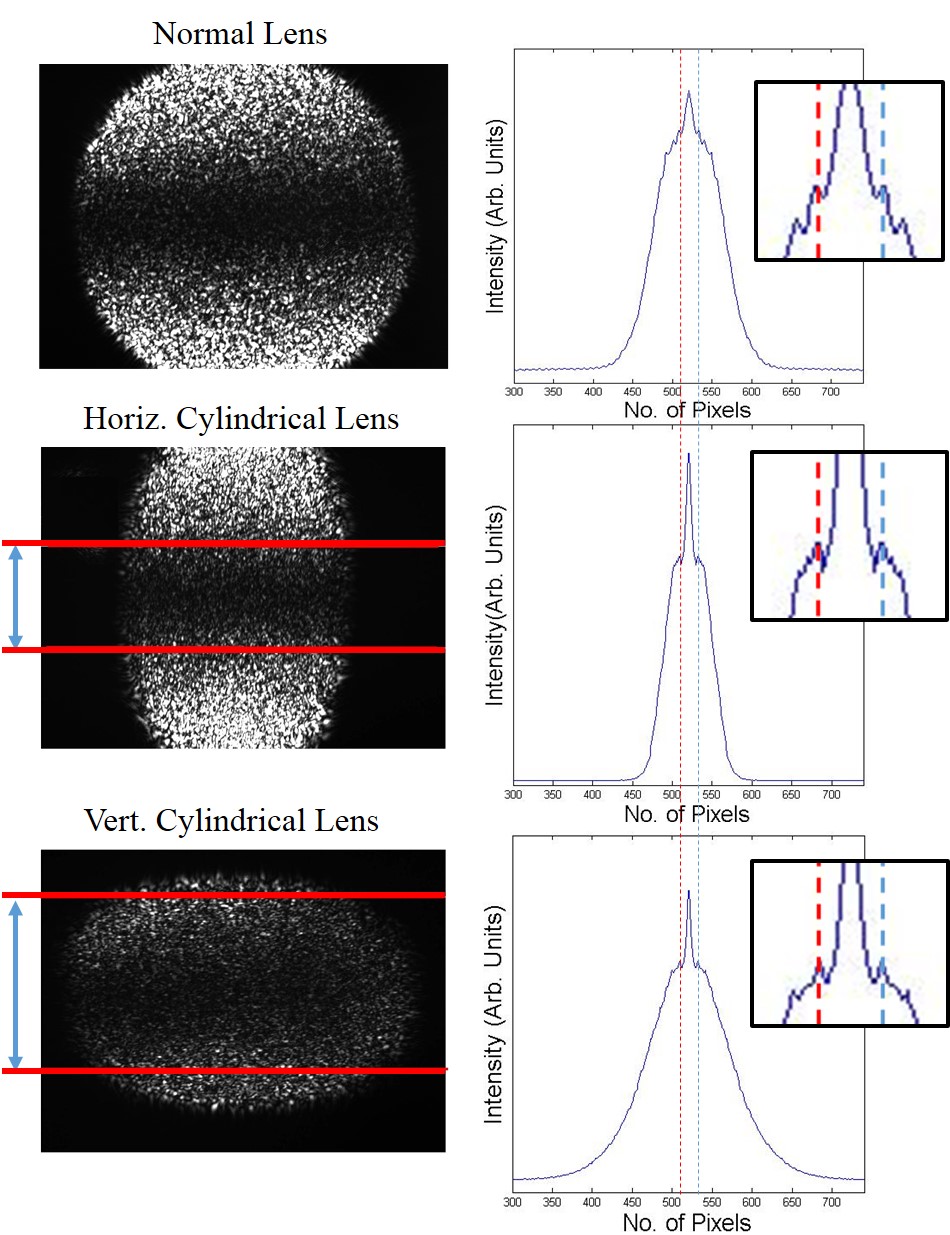}}
\caption{SHEL of pseudo-thermal light in different cases. The first row shows the SHEL observed using normal lens to focus the pseudo-thermal light and the second is using vertical cylindrical lens while the third is using horizontal. The autocorrelations are listed in the second column. It can be seen that pairs of peaks emerge in each picture, but the positions are the same. This can be interpreted as that the SHEL of partially coherent beam is a superposition of coherent speckles' SHEL. Autocorrelation reveals out the SHEL of unit speckles. The outermost pair of peaks can be accused as higher order autocorrelation peaks. (The frosted glass is rotating in this experiment. Autocorrelations are obtained by averaging 20000 results. Each picture's exposure time is 10ms. Gain is set to be 500. The size of pixel of the CCD we use is 6.45$\mu$m by 6.45$\mu$m.)}
\label{2}
\end{figure}

To explain this result, we undertake several experiments. Autocorrelation is commonly used to reveal the information of individual speckles when dealing with pseudo-thermal light. Same as the method used in \cite{Aiello_1}, the autocorrelation is obtained and shown in, especially for the central part of the dark strip, Fig. \ref{2}. The autocorrelation exhibits that aside from the general shape of Gaussian distribution, there are pairs of peaks spreading symmetrically about the center. One reasonable interpretation is that the peaks are the SHEL of coherent units which form statistically the SHEL of pseudo-thermal beam. We may regard this as our preliminary guess.

To verify our preliminary guess of the formation of pseudo-thermal SHEL is right, we apply the cylindrical lens to the pseudo-thermal light. Here we take two steps to explain the function of cylindrical lens. In the first step, we need to notice that cylindrical lens can change the overlapping density of speckles. Cylindrical lens limits the dimension of the total light beam in one direction. Given that the degree of spatial coherence is fixed, the total number of speckles is also a constant. So, by limiting the dimension of the beam along one specific dimension, the overlapping density of speckles will increase along that direction. The second step is to clarify the effect of the overlapping density to the width of the dark strip. If the SHEL of pseudo-thermal light is built by the SHEL of coherent speckles, the residue mottle in the dark strip should be due to the incoherence between single speckles. Now, it is natural to think that if increasing the overlapping density, more residues will be generated in the dark strip, i.e. the width of the dark strip will drop.

Let's summarize the effect of cylindrical lens supposing our guess is right. We use the setup as shown in Fig. \ref{1}(b) to adjust the overlapping density of speckles. If the speckles were converged tightly in the vertical direction onto the prism, then the strip, which is horizontal for SHEL, should be relatively narrow. On the other hand, if converged loosely, the strip should be relatively broad since the speckles spread in a larger area. By switching the cylindrical lens to horizontal/vertical position, the converging density of the speckles can be changed. We can anticipate that if our preliminary conclusion is correct, the strip would be relative narrow if the lens is horizontal, while the strip would be broad if the lens is vertical.

The results captured by the CCD matches our prediction, which are shown in Fig. \ref{2}. This demonstrates that the whole picture of SHEL of pseudo-thermal light is composed of large amount of SHEL of speckles. The red lines in Fig. \ref{2} illustrate the positions where the average light intensities equal. When the cylindrical lens is set horizontal, the strip in the center of the intensity intersection is narrower than the scenario when set vertical. However, the positions of the peaks (see insets) in the autocorrelation remain the same. The reason for this result is that at the same incident angle, every speckle act as its own. For speckles, which are of similar size and incidents at the same angle, the spatial shifts in SHEL of each speckle are more or less the same. Thus the linear polarized portion in the central part of every speckle have the same width. Then filtered by the polarizer (key step of weak measurement), every speckle appears a dark strip of the same width and this width is invariant with overlapping density. By overlapping big amount of these speckles, we get the SHEL of pseudo-thermal light.

\begin{figure}[htbp]
\centerline{\includegraphics[width=.8\columnwidth]{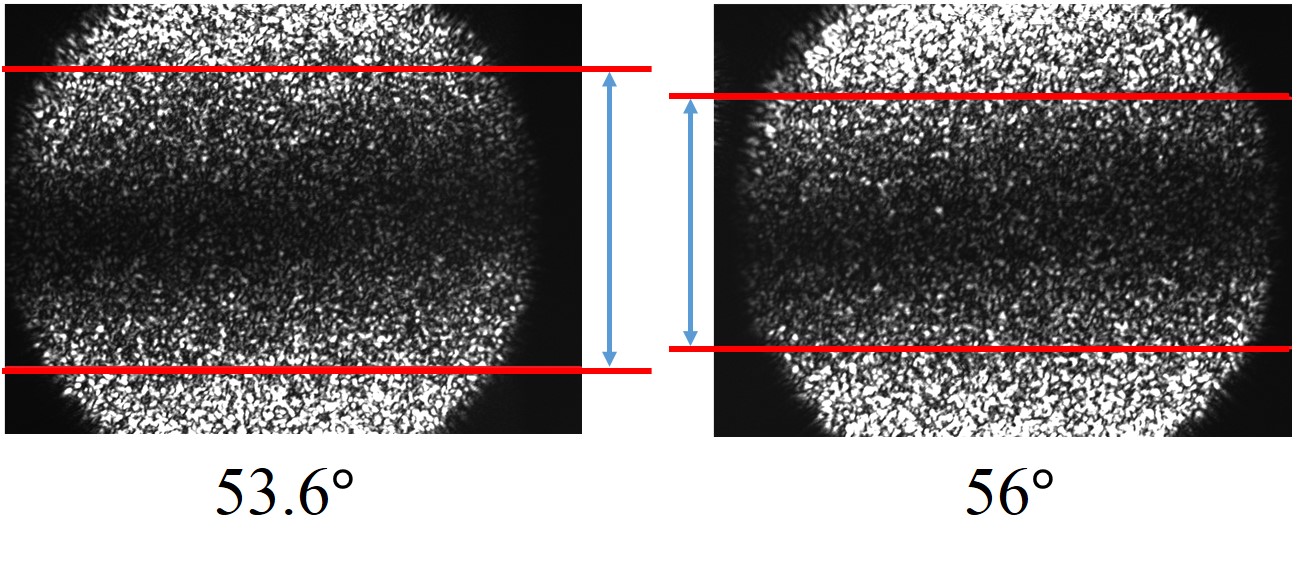}}
\caption{SHEL of partially coherent beams of different incident angles. Coherent speckles are confirmed that the SHEL will decline if the incident angle approaching to the Brewster angle ($56.55^{\circ}$ for BK7 prism) \cite{Enhanced}. This experiment proves that partially coherent beam follows the same trend.}
\label{3}
\end{figure}

To further prove our interpretation is correct, we test it at different incident angles. People have revealed that the SHEL of coherent beams decreases as the incident angle approaches to the Brewster angle ($56.55^{\circ}$ here)\cite{Enhanced}. Assuming our interpretation is right for the pseudo-thermal light, the SHEL of PCB should also decease when the incident angle is set close to the Brewster angle. By setting the pseudo-thermal light incident at different angles, we obtain a general trend of how the width of the dark strip in the pseudo-thermal beam varies. We track the strip width trend of the whole beam rather than autocorrelation results since we are limited by the CCD resolution. In our experiment, one speckle occupies an area of about 20 pixels by 20 pixels on the CCD camera and it is separated to two parts by the dark strip. It is almost impossible to resolve each part's center (maximum intensity point), since each part only has around 6 pixels in length. In Fig. \ref{3}, two typical pictures captured at different incident angles ($53.6^{\circ}$ and $56^{\circ}$) are given. As before, the red line in the pictures represent the same average light intensity. We can see that by altering the incident angle approaching to the Brewster angle, the width of the strip decreases, in other words, the SHEL is smaller. This matches the previous work done for coherent light. Therefore, our interpretation is legitimate to explain the SHEL of pseudo-thermal light.

\begin{figure}[htbp]
\centerline{\includegraphics[width=1\columnwidth]{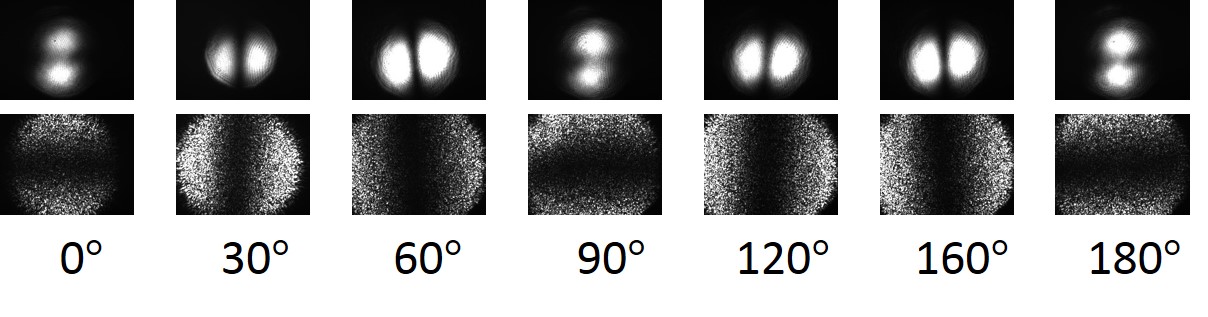}}
\caption{Rotation of the strip of partially coherent beams of different polarization accords to the rotation of coherent beams. We can determine that our theory apply to both Imbert-Fedorov (IF) shift and Goos-H\"anchen Shift.}
\label{4}
\end{figure}

It also worth noticing that, not limited in the SHEL scenario, our interpretation can also be applied to the Goos-H\"anchen (GH) shift. By using incident beams with different polarization, the case when SHEL and GH shift both happen can be cleared, as given in Fig. \ref{4}. The first row illustrates the cases for highly coherent beams with different polarization and the second is for the pseudo-thermal beams. The angles listed are the polarizer's angles relative to the initial position. It is shown that the directions of the strip for both the coherent and pseudo-thermal beams are synchronous. This synchronization means that regardless of what type of shifts (GH/IF) that happens, the explanation of the speckle's shifts forming the whole beam's shift is feasible.

\begin{figure}[htbp]
\centerline{\includegraphics[width=.8\columnwidth]{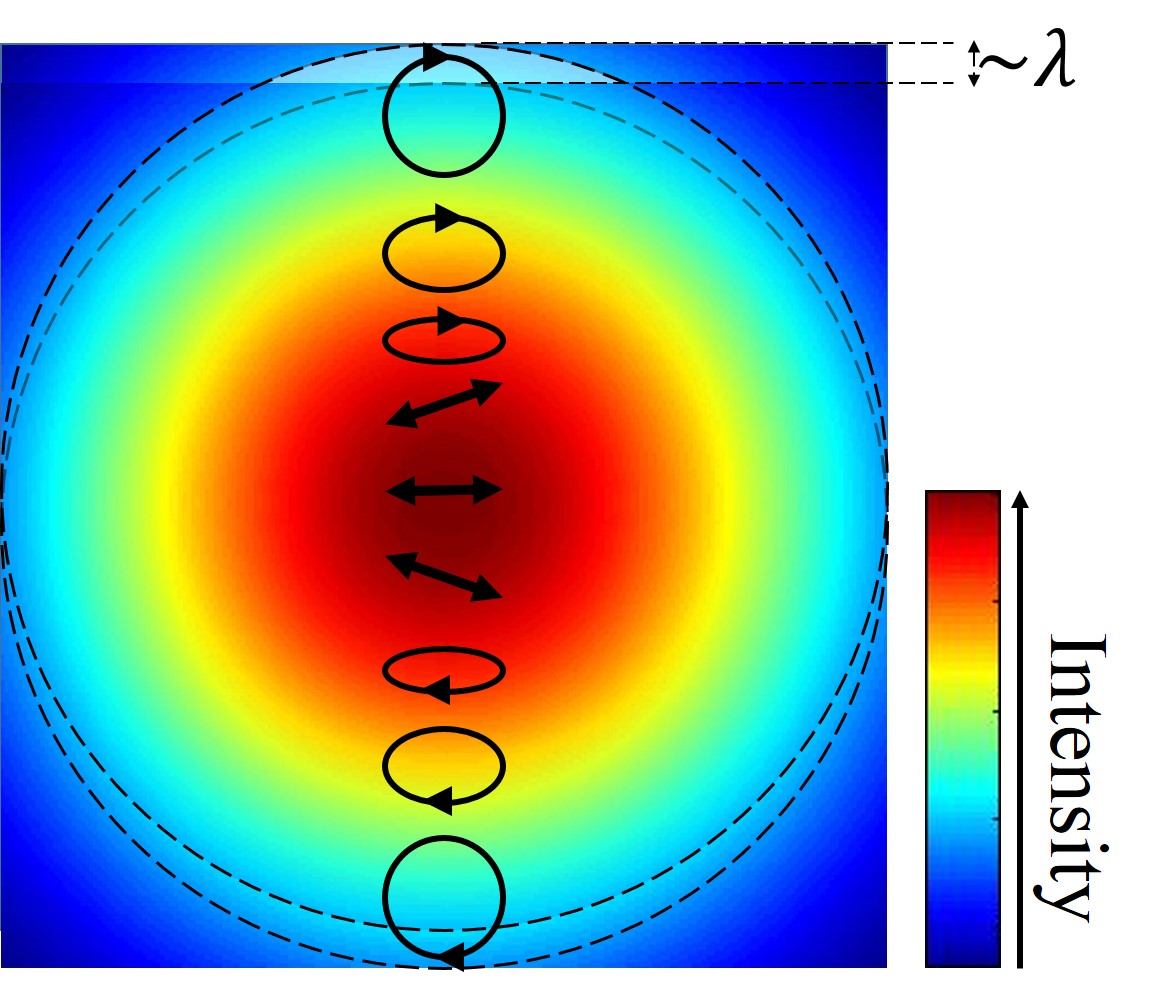}}
\caption{Sketch of intensity distribution and polarization property of coherent unit speckles with SHEL. The right/left circular polarization part of the incident beam is shifted due to the IF effect. The separation of two polarization components is of an order of wavelength. Because the intensity follows a Gaussian-like distribution, the overall polarization is illustrated as the black arrows. In fact, this sketch is also true for pseudo-thermal beam SHEL.}
\label{5}
\end{figure}

Till now, our experiments results indicate that the formation of the SHEL of pseudo-thermal light is due to the large collection of SHEL of small speckles, but we still lack a proper explication of the cause of the dark strip. As a matter of fact, the origin of the dark strip of the pseudo-thermal profile is the coherence remained in the pseudo-thermal light. A sketch is given as Fig. \ref{5}. The average intensity profile of beam speckles observes the Gaussian distribution, and the polarization properties are shown as the black arrows. On the other hand, the pseudo-thermal light as a whole abides by the Super-Poisson distribution. Considering the Super-Poisson distribution is also Poisson-like (high in the center, decline outward) and the pseudo-thermal beam is at least partial coherent, therefore the general intensity and polarization of pseudo-thermal light can also be described by Fig. \ref{5} roughly. Furthermore, both the speckles and partial coherent beam have dark strip appeared after weak measurement. Besides, the strip of the pseudo-thermal light isn't purely dark can be accredited to the incoherence of the pseudo-thermal light. If the beam is totally coherent, the strip will be purely dark, just like the speckles. 
\balance However, if the beam is totally incoherent, then the dark strip will not appear at all. It will be like our expectation that SHEL can only make the profile dim homogeneously. The real case of pseudo-thermal light is a scenario between fully coherent beam and total incoherent beam. Thus, the SHEL of pseudo-thermal light has a strip in the center, but it's not purely dark.

In summary, we propose an unsolved phenomenon observed when weak measurement and pseudo-thermal light are utilized in the experiment of SHEL. Similar to the SHEL of coherent light, the SHEL of pseudo-thermal light also shows a dark strip when through the technic weak measurement to measure. But the strip of pseudo-thermal light has scattered speckles in it. To figure out the reason, we perform several experiments. The autocorrelation of the central part of the strip is firstly retrieved and pairs of peaks emerge on top of the general Gaussian intensity distribution. We make the preliminary guess that the SHEL of pseudo-thermal light is composed of the SHEL of coherent speckles statistically. Then two more tests are conducted, including changing the overlapping density of speckles, and changing the incident angles of the beam. All the results can be properly anticipated and explained by our interpretation. Finally, the appearance of the strip is also explainable when the shape of the intensity distribution is taken into account. Hence, we make the conclusion that because of the coherence of speckles and pseudo-thermal beam and the Gaussian-like intensity distributions, coherent speckles and partially coherent beams both have dark strip when weak measurement is applied. However, there exists difference that, thanks to the incoherence of the pseudo-thermal light, there are fares remained in the dark strip, which makes the reveal of the SHEL of unit speckles through autocorrelation possible.

Nevertheless, we admit that there is still a lot work not done yet. Our interpretation only gives a qualitative explanation, lacking a quantitative analysis. Due to the limit of the CCD resolution, we are unable to reveal the variance of SHEL of single speckles. With a CCD with higher resolution, measurement of pseudo-thermal light combining weak measurement and autocorrelation could be developed to a new method of measurement, which may give a better accuracy.

We acknowledge financial support from the National Natural Science Foundation of China (NSFC) under grants 11204235 and 11374238, the Fundamental Research Funds for the Central Universities under grant xjj2014097, and the Doctoral Fund of the Ministry of Education of China under Grant 20120201110035.


\pagebreak

\bibliographystyle{osajnl}
\bibliography{SHEL} 

\begin{thebibliography}{10}
\newcommand{\enquote}[1]{``#1''}

\bibitem{fedorov1955to}
F.~I. Fedorov, \enquote{To the theory of total reflection,} Doklady Akademii
  Nauk SSSR \textbf{105}, 465--8 (1955).

\bibitem{Overview}
K.~Y. Bliokh and A.~Aiello, \enquote{Goos--h{\"a}nchen and imbert--fedorov beam
  shifts: an overview,} Journal of Optics \textbf{15}, 014001 (2013).

\bibitem{onoda2004hall}
M.~Onoda, S.~Murakami, and N.~Nagaosa, \enquote{Hall effect of light,} Physical
  review letters \textbf{93}, 083901 (2004).

\bibitem{goos1955TEin}
F.~Goos and H.~H{\"a}nchen, \enquote{Tein neuer und fundamentaler versuch zur
  totalreflexion,} Annalen der Physik \textbf{436}, 333–346 (1947).

\bibitem{jayaswal2014observation}
G.~Jayaswal, G.~Mistura, and M.~Merano, \enquote{Observation of the
  imbert--fedorov effect via weak value amplification,} Optics letters
  \textbf{39}, 2266--2269 (2014).

\bibitem{jayaswal2013weak}
G.~Jayaswal, G.~Mistura, and M.~Merano, \enquote{Weak measurement of the
  goos--h{\"a}nchen shift,} Optics letters \textbf{38}, 1232--1234 (2013).

\bibitem{wang2013goos}
L.-G. Wang, S.-Y. Zhu, and M.~S. Zubairy, \enquote{Goos-h{\"a}nchen shifts of
  partially coherent light fields,} Physical review letters \textbf{111},
  223901 (2013).

\bibitem{chuang2015negative}
Ziauddin, Y.-L. Chuang, and R.-K. Lee, \enquote{Negative and positive
  goos-h\"anchen shifts of partially coherent light fields,} Phys. Rev. A
  \textbf{91}, 013803 (2015).

\bibitem{rodriguez2010optical}
O.~G. Rodr{\'\i}guez-Herrera, D.~Lara, K.~Y. Bliokh, E.~A. Ostrovskaya, and
  C.~Dainty, \enquote{Optical nanoprobing via spin-orbit interaction of light,}
  Physical review letters \textbf{104}, 253601 (2010).

\bibitem{Ziauddin2012control}
Ziauddin and S.~Qamar, \enquote{Control of the goos-h\"anchen shift using a
  duplicated two-level atomic medium,} Phys. Rev. A \textbf{85}, 055804 (2012).

\bibitem{haefner2009spin}
D.~Haefner, S.~Sukhov, and A.~Dogariu, \enquote{Spin hall effect of light in
  spherical geometry,} Physical review letters \textbf{102}, 123903 (2009).

\bibitem{korger2014observation}
J.~Korger, A.~Aiello, V.~Chille, P.~Banzer, C.~Wittmann, N.~Lindlein,
  C.~Marquardt, and G.~Leuchs, \enquote{Observation of the geometric spin hall
  effect of light,} Physical review letters \textbf{112}, 113902 (2014).

\bibitem{lv2012spin}
Y.~Lv, Z.~Wang, Y.~Jin, M.~Cao, L.~Han, P.~Zhang, H.~Li, H.~Gao, and F.~Li,
  \enquote{Spin polarization separation of light reflected at brewster angle,}
  Optics letters \textbf{37}, 984--986 (2012).

\bibitem{Science}
O.~Hosten and P.~Kwiat, \enquote{Observation of the spin hall effect of light
  via weak measurements,} Science \textbf{319}, 787--790 (2008).

\bibitem{dressel2014colloquium}
J.~Dressel, M.~Malik, F.~M. Miatto, A.~N. Jordan, and R.~W. Boyd,
  \enquote{Colloquium: Understanding quantum weak values: Basics and
  applications,} Reviews of Modern Physics \textbf{86}, 307 (2014).

\bibitem{luo2009spin}
H.~Luo, S.~Wen, W.~Shu, Z.~Tang, Y.~Zou, and D.~Fan, \enquote{Spin hall effect
  of a light beam in left-handed materials,} Physical Review A \textbf{80},
  043810 (2009).

\bibitem{nalitov2015spin}
A.~Nalitov, G.~Malpuech, H.~Ter{\c{c}}as, and D.~Solnyshkov,
  \enquote{Spin-orbit coupling and the optical spin hall effect in photonic
  graphene,} Physical Review Letters \textbf{114}, 026803 (2015).

\bibitem{Aiello_0}
A.~Aiello and J.~Woerdman, \enquote{Role of spatial coherence in
  goos-h{\"a}nchen and imbert--fedorov shifts,} Optics letters \textbf{36},
  3151--3153 (2011).

\bibitem{Aiello_1}
W.~L{\"o}ffler, A.~Aiello, and J.~Woerdman, \enquote{Spatial coherence and
  optical beam shifts,} Physical review letters \textbf{109}, 213901 (2012).

\bibitem{merano2012observation}
M.~Merano, G.~Umbriaco, and G.~Mistura, \enquote{Observation of nonspecular
  effects for gaussian schell-model light beams,} Physical Review A
  \textbf{86}, 033842 (2012).

\bibitem{Measurement}
Y.~Qin, Y.~Li, H.~He, and Q.~Gong, \enquote{Measurement of spin hall effect of
  reflected light,} Optics letters \textbf{34}, 2551--2553 (2009).

\bibitem{Enhanced}
H.~Luo, X.~Zhou, W.~Shu, S.~Wen, and D.~Fan, \enquote{Enhanced and switchable
  spin hall effect of light near the brewster angle on reflection,} Physical
  Review A \textbf{84}, 043806 (2011).

\end{thebibliography}


\begin{thebibliography}{99}
\bibitem{fedorov1955to}F. I. Fedorov, Dokl. Akad. Nauk SSSR \textbf{105}, 465 (1955).
\bibitem{Overview}K. Y. Bliokh and A. Aiello, J. Opt. \textbf{15}, 014001 (2013).
\bibitem{}M. Onoda, S. Murakami, and N. Nagaosa, Phys. Rev. Lett. \textbf{93}, 083901 (2004).
\bibitem{fgoos1955TEin}F. Goos and H. H{\"a}nchen, Ann. Phys. \textbf{436}, 333 (1947).
\bibitem{Jayaswal}G. Jayaswal, G. Mistura, and M. Merano, Opt. Lett. \textbf{39}, 2266 (2014).
\bibitem{}G. Jayaswal, G. Mistura, and M. Merano, Opt. Lett. \textbf{38}, 1232 (2013).
\bibitem{Wang}L.-G. Wang, S.-Y. Zhu, and M. S. Zubairy, Phys. Rev. Lett. \textbf{111}, 213901 (2013).
\bibitem{Chuang}Ziauddin, Y.-L. Chuang, and R.-K. Lee, Phys. Rev. A \textbf{91}, 013803 (2015).
\bibitem{}O. G. Rodr{\'\i}guez-Herrera, D. Lara, K. Y. Bliokh, E. A. Ostrovskaya, and C. Dainty, Phys. Rev. Lett. \textbf{104}, 253601 (2010).
\bibitem{}Ziauddin and S. Qamar, Phys. Rev. A \textbf{85}, 055804 (2012).
\bibitem{}D. Haefner, S. Sukhov, and A. Dogariu, Phys. Rev. Lett. \textbf{102}, 123903 (2009).
\bibitem{}J. Korger, A. Aiello, V. Chille, P. Banzer, C. Wittmann, N. Lindlein, C. Marquardt, and G. Leuchs, Phys. Rev. Lett. \textbf{112}, 113902 (2014).
\bibitem{}Y. Lv, Z. Wang, Y. Jin, M. Cao, L. Han, P. Zhang, H. Li, H. Gao, and F. Li, Opt. Lett. \textbf{37}, 984 (2012).
\bibitem{Science}O. Hosten and P. Kwiat, Science \textbf{319}, 787 (2008).
\bibitem{}J. Dressel, M. Malik, F. M. Miatto, A. N. Jordan, and R. W. Boyd, Rev. Mod. Phys. \textbf{86}, 307 (2014).
\bibitem{}H. Luo, S. Wen, W. Shu, Z. Tang, Y. Zou, and D. Fan, Phys. Rev. A \textbf{80}, 043810 (2009).
\bibitem{}A. Nalitov, G. Malpuech, H. Terc ̧as, and D. Solnyshkov, Phys. Rev. Lett. \textbf{114}, 026803 (2015).
\bibitem{Aiello_0}A. Aiello and J. Woerdman, Opt. Lett. \textbf{36}, 3151 (2011).
\bibitem{Aiello_1}W. L{\"o}ffler, A. Aiello, and J. Woerdman, Phys. Rev. Lett. \textbf{109}, 213901 (2012).
\bibitem{}M. Merano, G. Umbriaco, and G. Mistura, Phys. Rev. A \textbf{86}, 033842 (2012).
\bibitem{Measurement}Y. Qin, Y. Li, H. He, and Q. Gong, Opt. Lett. \textbf{34}, 2551 (2009).
\bibitem{Enhanced}H. Luo, X. Zhou, W. Shu, S. Wen, and D. Fan, Phys. Rev. A \textbf{84}, 043806 (2011).

\end{thebibliography}

\begin{thebibliography}{99}
\bibitem{Overview} K. Y. Bliokh and A. Aiello, "Goos--H{\"a}nchen and Imbert--Fedorov beam shifts: an overview," Journal of Optics \textbf{15}, 014001 (2013).
\bibitem{Aiello_0}A. Aiello and J. Woerdman, "Role of spatial coherence in Goos-H{\"a}nchen and Imbert--Fedorov shifts," Optics letters \textbf{36}, 3151?3153 (2011).
\bibitem{Aiello_1}W. L{\"o}ffler, A. Aiello, and J. Woerdman, "Spatial Coherence and Optical Beam Shifts," Physical review letters \textbf{109}, 213901 (2012).
\bibitem{Science}O. Hosten and P. Kwiat, "Observation of the spin Hall effect of light via weak measurements," Science \textbf{319}, 787? 790 (2008).
\bibitem{Measurement}Y. Qin, Y. Li, H. He, and Q. Gong, "Measurement of spin Hall effect of reflected light," Optics letters \textbf{34}, 2551?2553 (2009).
\bibitem{Enhanced}H. Luo, X. Zhou, W. Shu, S. Wen, and D. Fan, "Enhanced and switchable spin Hall effect of light near the Brewster angle on reflection," Physical Review A \textbf{84}, 043806 (2011).
\bibitem{jayaswal2014observation}G. Jayaswal, G. Mistura, and M. Merano, "Observation of the Imbert–Fedorov effect via weak value amplification," Optics letters, \textbf{39}, 2266-2269 (2014).
\bibitem{wang2013goos}L. G. Wang, S. Y. Zhu, and M. S. Zubairy, "Goos--H{\"a}nchen Shifts of Partially Coherent Light Fields," Physical review letters, \textbf{111}, 223901 (2013).
\bibitem{chuang2015negative}Y. L. Chuang, R. K. Lee, "Negative and positive Goos--H{\"a}nchen shifts of partially coherent light fields," Physical Review A, \textbf{91}, 013803 (2015).
\bibitem{luo2009spin}H. Luo, S. Wen, W. Shu, Z. Tang, Y. Zou, and D. Fan, "Spin Hall effect of a light beam in left-handed materials," Physical Review A, \textbf{80}, 043810 (2009).

\end{thebibliography}

\end{document}